# Rate Splitting in VCSEL-based Optical Wireless Networks

Khulood Alazwary, Ahmad Adnan Qidan, Taisir El-Gorashi and Jaafar M. H. Elmirghani
*School of Electronic and Electrical Engineering University of Leeds, Leeds, United Kingdom*
{elkal, A.A.Qidan, t.e.h.elgorashi, J.M.H.Elmirghani}@leeds.ac.uk

*Abstract*— Optical wireless communication is an effective potential solution for enabling high speed next generation cellular networks. In this paper, laser sources, in particular, vertical-cavity surface-emitting (VCSEL) lasers are used for data transmission due to their high modulation speed compared with light emitting diode (LED) sources. To manage multi-user interference, rate splitting (RS) is implemented where the message of a user is split into common and private parts. However, the performance of RS is limited in high density networks. Therefore, hierarchical rate splitting (HRS) particularly suited in high density networks is considered. The results demonstrate the high data rate achieved by using VCSELs. Moreover, HRS is more suitable for achieving high performance in optical networks compared with RS.

## I. INTRODUCTION

The demands for connecting a massive number of devices to the Internet through Internet of things (IoT) as well as single high data rate terminals represent major challenges for current cellular networks. This requires a paradigm shift towards innovative cellular technologies that can support high data rates. Optical wireless communication is a potential solution that can support and meet the enormous growing demand for high data rates due to its high capacity [1]–[5]. In visible light communication (VLC), data rates can reach up to 20 Gbps with low infrastructure cost [3], [6]. Therefore, optical wireless systems have gained increased interest and are considered for sixth generation (6G) communication systems. This has motivated the investigation of different adaptation techniques such as beam power, beam angle, and beam delay adaptations in order to improve the capacity of optical communication links [4]–[15]. However, energy efficiency is an area that needs more attention [16], [17]. In this context, VCSELs can provide focused beams that can be adapted, for data transmission more efficiently than light emitting diodes (LED) due to their high modulation speed, spectral purity and low cost [18]. To further enhance the capacity of optical wireless systems, on the users side, angle diversity receivers (ADRs) are proposed to tackle directed and undirected interference [19]–[25].

In multi-user optical cellular systems, several optical orthogonal multiple access (OMA) schemes have been proposed and investigated for interference management such as time division multiple access (TDMA), orthogonal frequency division multiple access (OFDMA), optical code division multiple access (OCDMA) [26], wavelength division multiple access (WDMA) [27], [28], and space division multiple access (SDMA). In addition, non orthogonal multiple access (NOMA) has been recently introduced as a spectrum-efficient multiple access scheme where multiple users can be served simultaneously over the same resources , i.e., time and frequency slots [29].

Recently, rate splitting (RS) schemes have been proposed in multiple-input single-output (MISO) RF systems as a promising transmission scheme, which is capable of providing high data rates compared to other traditional multiple access schemes. Besides, RS is more efficient in managing the interference among users even in the absence of accurate channel state information (CSI) at transmitters, and therefore, it is characterized by its high spectral and energy efficiency [30]. However, the performance of RS is severely limited in serving a large number of users. Therefore, hierarchical rate splitting (HRS) has been particularly proposed for massive multiple-input and multiple-output (MIMO) RF systems to enhance the sum rate and alleviate CSI requirements [31].

In this paper, the performance of two transmission schemes, RS and HRS, is evaluated in a VCSEL-based optical wireless network. Therefore, identifying the impact of increasing users on the achievable user rate of RS and HRS schemes. VCSELs are used as transmitter units to provide high data rate communication. To the best of our knowledge, this is the first work that investigates the efficient integration of VCSEL and RS in optical wireless systems. The results demonstrate the high achievable rate in the VCSEL-RS optical wireless system. Moreover, HRS provides high data rates compared to RS where users are divided into multiple groups managing the interference efficiently among them.

The rest of this paper is organized as follows: Section II describes the system configuration, including the room, VCSEL and ADR configuration, while the simulation results are given and discussed in Section III and the conclusions are stated in Section IV.

## II. SYSTEM MODEL

In this work, we considered an empty room with dimensions (length × width × height) of 5 m × 5 m × 3 m. In this room, four optical transmitters are distributed on the ceiling as shown in Fig. 1. Each optical transmitter is composed of multiple VCSELs. The transmitted power of the VCSEL transmitter is based on the beam width, $W_0$, the wavelength and the distance through which the beam travels.

An ADR that consists of four optical photodiode detectors similar to [27] is used as shown in Fig. 2. Each optical photodetector in the ADR has a narrow Field of View (FOV), and its direction is given by the Azimuth ($Az$) and the Elevation ($El$) angles collecting signals from different sectors of the room.

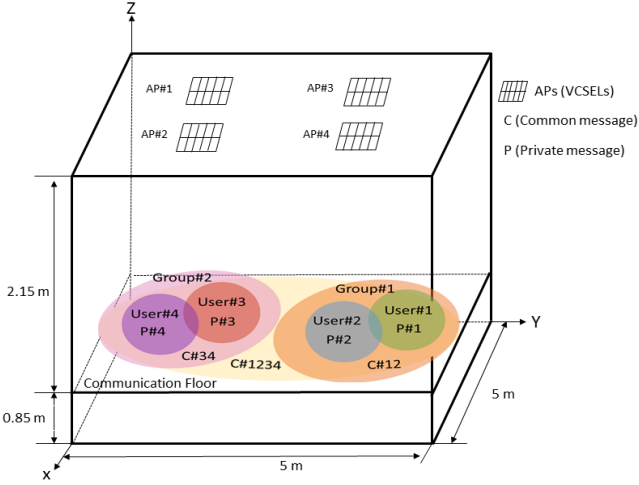

Fig 1. Room configuration and HRS illustrative example.

Twenty users are randomly distributed over the communication floor, which is a plane 0.85 m above the ground, forming spatial clusters. In this work, we consider a scenario where each user has a line of sight link with each transmitter, and the diffuse component of the channel is neglected for the sake of simplicity. **Error! Reference source not found.** shows the overall simulation parameters of the room, VCSEL and ADR detector.

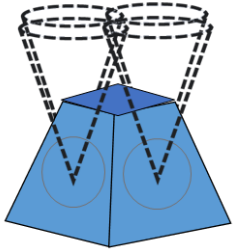

Fig 2. Angle Diversity Receiver

### III. IMPLEMENTING RATE SPLITTING

In this paper we deploy and analyse the performance of RS and HRS schemes in downlink MISO broadcast channel (MISO BC) using VCSEL transmitters.

#### A. Rate Splitting

In RS scheme, the message intended to a user is split into common and private messages. The common message is defined as public codebook on which all private messages of users are superimposed. Then, linear precoding is considered for transmission. In particular, the transmitted signal to $K$ users can be expressed as:

$$X = \sqrt{P_c}\mathbf{w}_c s_c + \sum_{k=1}^{K} \sqrt{P_k}\mathbf{w}_k s_k, \forall k \in K, \quad (1)$$

where $\mathbf{s} = [s_1, s_2, \ldots s_k]$ is the data vector intended for $K$ users and $s_c$ is the common message. Moreover, $\mathbf{W} = [\mathbf{w}_1, \mathbf{w}_2 \ldots \ldots, \mathbf{w}_k]$ is designed according to zero forcing precoder, and $\mathbf{w}_c$ is the unit-norm precoding vector of the common message. Furthermore, the power allocated to each private message is given by $P_k = Pt/K$ where $t \in (0,1]$ is the fraction of the total power $P$ that is assigned to the private messages. The remaining power is allocated to the common message $P_c = P(1-t)$. At this point, the received signal of user $k$ is given by

$$y_k = \sqrt{P_k}\mathbf{h}_k^T\mathbf{w}_k s_k + \sum_{j \neq k}^{K} \sqrt{P_j}\mathbf{h}_k^T\mathbf{w}_j s_j + z_k, \quad (2)$$

where $\mathbf{h}_k^T$ is the channel gain vector between user $k$ and the optical transmitters and $z_k$ is real valued additive white Gaussian noise (AWGN) with zero mean and variance $\sigma_k^2$ representing thermal and shot noise.

Table I. System Parameters

| Parameters | Configurations | | | |
|---|---|---|---|---|
| **VCSEL** | | | | |
| Beam waist, $W_0$ | 5 $\mu m$ | | | |
| Number of VCSELs per optical transmitter unit | 10 | | | |
| VCSEL wavelength | 850 $nm$ | | | |
| Semi-angle of reflection element at half power | 15° | | | |
| **Room** | | | | |
| Width × Length × Height (x, y, z) | 5m × 5 m × 3m | | | |
| Number of optical transmitter units | 8 | | | |
| Optical transmitter locations (x, y, z) | (3.5m, 3.5m, 3m), (1.5m, 3.5m, 3m), (3.5 m, 1.5m, 3m), (1.5 m, 1.5 m, 3m) | | | |
| **Angle Diversity Receiver** | | | | |
| Responsivity | 0.4 A/W | | | |
| Number of photodetectors | 4 | | | |
| Area of the photodetector | 20 mm$^2$ | | | |
| Receiver noise current spectral density | 4.47 pA/√Hz | | | |
| Receiver bandwidth | 5 GHz | | | |
| Photodetector | *1* | *2* | *3* | *4* |
| Azimuth angles | 0° | 90° | 180° | 270 |
| Elevation angles | 60° | 60° | 60° | 60° |
| Field of view (FOV) | 25° | 25° | 25° | 25° |

To decode the desired signal, each user follows two steps to decode its message: the user firstly decodes the common stream by treating all private streams as noise. Secondly, the user decodes its private stream after removing the decoded common stream using Successive Interference Cancellation (SIC). Therefore, the signal to interference and noise ratio (SINR) of the common message, $\gamma_k^c$, and the SINR of the private message, $\gamma_k^p$, transmitted to user $k$ can be defined as

$$\gamma_k^c = \frac{P_c|\mathbf{h}_k^T\mathbf{w}_c|^2}{\sum_{j=1}^{K} P_j|\mathbf{h}_k^T\mathbf{w}_j|^2 + \sigma_k^2} \quad (3)$$

$$\gamma_k^p = \frac{P_k |\mathbf{h}_k^T \mathbf{w}_k|^2}{\sum_{j \neq k} P_j |\mathbf{h}_k^T \mathbf{w}_j|^2 + \sigma_k^2} \quad (4)$$

For the proposed RS, the achievable rate of the common message is given as

$$R_c^{RS} = \log_2(1 + \gamma^c), \quad (5)$$

where $\gamma^c = \min_k \{\gamma_k^c\}$ to ensure that the common message is successfully decoded by all users. The achievable rate of the private message at user $k$ is $R_k^{RS} = \log_2(1 + \gamma_k^p)$. Thus the sum rate of the private messages is

$$R_p^{RS} = \sum_{k=1}^K R_k^{RS} = \sum_{k=1}^K \log_2(1 + \gamma_k^p). \quad (6)$$

Therefore, the sum rate of RS is given as

$$R_{\text{sum}}^{RS} = R_c^{RS} + R_p^{RS} \quad (7)$$

## B. Hierarchical Rate Splitting (HRS)

HRS with a two-tier precoder has been proposed to enhance the sum rate and alleviate CSIT requirement as well as the effect of large number of users by exploiting the knowledge of spatial correlation matrix at transmitters. To apply HRS, $K$ users are divided into multiple groups denoted as $G$. Each group contains a set of users grouped based on the minimum distance among them. After that, two RS schemes referred to as outer and inner RS are implemented for managing inter-group interference and intra-group interference, respectively. The outer RS transmits one outer common message that can be decoded by all users aligning the inter-group interference. on the other hand, the inner RS transmits inner common messages that can decoded by subsets of users to mitigate the intra-group interference. The outer and inner common messages are superimposed over the private messages by using two-tier precoder. Therefore, the transmitted signal of HRS is given by

$$\mathbf{x} = \sum_{g=1}^G \mathbf{B}_g \mathbf{W}_g \mathbf{P}_g \mathbf{s}_g \quad (8)$$

where $\mathbf{B}_g$ is the outer precoder and $\mathbf{W}_g$ is the inner precoder. Moreover, $\mathbf{s}_g$ are messages transmitted to group $g$ and $\mathbf{P}_g$ is the power allocated to that group. The power allocated to each message is jointly determined by $\alpha$ and $\beta$, where $\beta \in (0,1]$ is the fraction of the total power allocated to the group messages, while $\alpha \in (0,1]$ is considered within the group as a fraction of power that is allocated to the private messages. That is, the power of the outer common message is $P_{oc} = P(1-\beta)$, the power of the inner common message is $P_{ic,g} = \frac{P\beta}{G}(1-\alpha)$ and the power of the private message is $P_{gk} = \frac{P\beta}{K}\alpha$. Then, the received signal at user $k$ in group $g$ is given by

$$y_{gk} = \sqrt{P_{gk}} \mathbf{h}_{gk}^T \mathbf{B}_g \mathbf{w}_{gk} s_{gk} + \sum_{j \neq k}^{K_g} \sqrt{P_{gj}} \mathbf{h}_{gk}^T \mathbf{B}_g \mathbf{w}_{gj} s_{gj} + \sum_{l \neq g}^G \mathbf{h}_{gk}^T \mathbf{B}_l \mathbf{W}_l \mathbf{P}_l \mathbf{s}_l + z_{gk}, \quad (9)$$

where $\sum_{g=1}^G K_g = K$. The decoding procedure of HRS starts by removing the outer and inner common messages from the received signal through performing SIC. Then, each user belonging to a group can decode its private message successfully. The Signal to Interference plus Noise Ratios of the outer and inner common messages and the private message of user $k$ can be expressed as

$$\gamma_{gk}^{oc} = \frac{P_{oc} |\mathbf{h}_{gk}^T \mathbf{w}_{oc}|^2}{\sum_{l=1}^G \sum_{j=1}^{K_g} P_{lj} |\mathbf{h}_{gk}^T \mathbf{B}_l \mathbf{w}_{lj}|^2 + \sum_{l=1}^G P_{ic,l} |\mathbf{h}_{gk}^T \mathbf{B}_l \mathbf{w}_{ic,l}|^2 + \sigma_{gk}^2} \quad (10)$$

$$\gamma_{gk}^{ic} = \frac{P_{ic,g} |\mathbf{h}_{gk}^T \mathbf{B}_g \mathbf{w}_{ic,g}|^2}{\sum_{l=1}^G \sum_{j=1}^{K_g} P_{lj} |\mathbf{h}_{gk}^T \mathbf{B}_l \mathbf{w}_{lj}|^2 + \sum_{l \neq g}^G P_{ic,l} |\mathbf{h}_{gk}^T \mathbf{B}_l \mathbf{w}_{ic,l}|^2 + \sigma_{gk}^2} \quad (11)$$

$$\gamma_{gk}^p = \frac{P_{gk} |\mathbf{h}_{gk}^T \mathbf{B}_g \mathbf{w}_{gk}|^2}{\sum_{l=1}^G \sum_{j \neq k}^{K_g} P_{lj} |\mathbf{h}_{gk}^T \mathbf{B}_l \mathbf{w}_{lj}|^2 + \sum_{l \neq g}^G P_{ic,l} |\mathbf{h}_{gk}^T \mathbf{B}_l \mathbf{w}_{ic,l}|^2 + \sigma_{gk}^2} \quad (12)$$

respectively. Moreover, the achievable rate of the outer common message is given by

$$R_{oc}^{HRS} = \log_2(1 + \gamma^{oc}), \gamma^{oc} = \min_{g,k} \{\gamma_{gk}^{oc}\}.$$

While, the sum rate of the inner common messages is given as $R_{ic}^{HRS} = \sum_{g=1}^G \log_2(1 + \gamma_g^{ic}), \gamma_g^{ic} = \min_k \{\gamma_k^{ic}\}$, and the sum rate of the private messages is given by

$R_p^{HRS} = \sum_{g=1}^G \sum_{k=1}^{K_g} \log_2(1 + \gamma_{gk}^p)$. Therefore, the sum rate of HRS is expressed as

$$R_{\text{sum}}^{HRS} = R_{oc}^{HRS} + R_{ic}^{HRS} + R_p^{HRS}. \quad (13)$$

For illustrative purposes, the methodology of HRS is described in Fig.1, considering the transmission to two groups, each with two users. The outer RS transmits one outer common stream (C#1234) for both groups. The inner RS transmits two inner common streams (C#12), (C#34), one inner common stream for each group. In this sense, each user follows three steps to decode its information. Focussing on user 1, the outer common stream (C#1234) is decoded treating the inner common streams (C#12,C#34) and private streams (P#1, P#2, P#3, P#4) as noise. Then, user 1 decodes its inner common stream by removing the decoded outer common stream using SIC and treating the private streams as noise. Finally, user 1 decodes its private streams after removing the decoded outer and inner common streams using SIC.

## IV. SIMULATION RESULTS

The simulation results in Fig. 3 and Fig. 4 present the achievable user rate of RS and HRS schemes. In Fig. 3, it can be seen that the achievable user rate decreases with the number of users due to interference subtraction as well as the lack of resources. Comparing RS and HRS, the performance of HRS is significantly better than RS regardless of the number of groups. This is because, HRS is based on exploiting the outer precoder and partitioning users into a set of non-interfering groups leading to less resultant interference. Notice that, the HRS scheme achieves better performance with $G = 5$ than $G = 10$ in all the considered scenarios. For instance, when $K = 10$, HRS with $G = 5$ achieves 8.7 Gbps user rate compared to 7.7 Gbps assuming $G = 10$. It is worth mentioning that increasing the number of groups results in reducing the complexity of interference management within each group. However, the noise increases as the size of groups decreases. Therefore, the trade-off between the size of groups and noise must be taken into consideration.

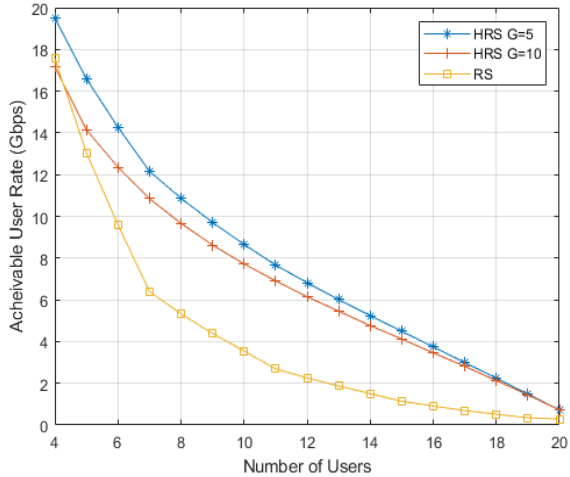

Fig 1. Achievable User Rate versus Number of users for RS and HRS.

In Fig. 4, the achievable user rate of RS and HRS is depicted against the beam waist $W_0$ of the VCSEL source. We considered and utilized VCSELs with a beam waist equal to 20 $\mu m$ as a starting point, then increased it by 10 $\mu m$ to a maximum value of 80 $\mu m$. It can be seen that the performance of HRS and RS improves with increase in the beam waist due to the fact that increasing the beam waist results in a high received power and less beam divergence. However, eye safety must be taken into consideration. In this sense, an optimal beam waist must be found to ensure achieving a high data rate without compromising eye safety. Finally, HRS is more suitable for our network achieving a higher user data rate than RS over different values of the beam waist.

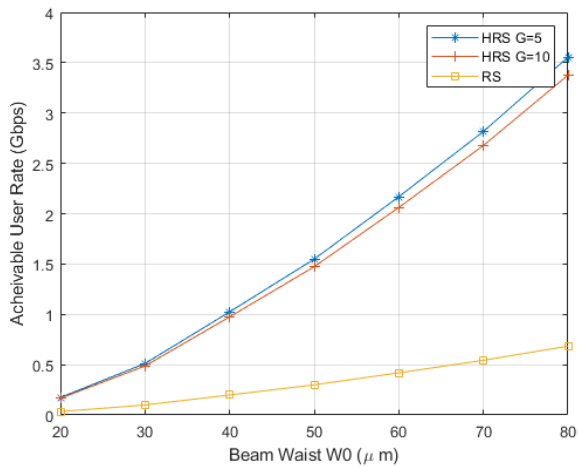

Fig 2. Achievable User Rate versus VCSEL Beam Waist for RS and HRS.

## V. CONCLUSIONS

In this paper, we evaluated the performance of RS and HRS in a VCSEL-based optical wireless network. First, we defined the system model, which is composed of multiple VCSELs serving multiple users. Then, the methodology of RS and HRS in managing the interference among multiple users is presented to derive the achievable rate. The simulation results demonstrate that HRS is more suitable for a high-density optical wireless network compared with RS. Moreover, the impact of the number of groups on the performance of HRS is shown. Finally, the achievable user rate for RS and HRS improves with increase in the the beam waist of VCSEL sources due to the associated increase in the received power. For future work, various optimization problems will be formulated for allocating the power efficiently among the messages of the users. In addition, consideration will be given to the enhancement of the transmitted power under eye safety constraints in MIMO VCSEL scenarios.

ACKNOWLEDGMENTS

The authors would like to acknowledge funding from the Engineering and Physical Sciences Research Council (EPSRC) INTERNET (EP/H040536/1), STAR (EP/K016873/1) and TOWS (EP/S016570/1) projects. KDA would like to thank King Abdulaziz University in the Kingdom of Saudi Arabia for funding her PhD scholarship. All data are provided in full in the results section of this paper.